\documentclass[aps,prd,reprint,longbibliography,titlepage,nofootinbib]{revtex4-2}

\usepackage{bm}
\usepackage[T1]{fontenc}
\usepackage{amsmath}
\usepackage{amsfonts}
\usepackage{mathtools} 
\usepackage{enumerate}
\usepackage{hyperref}
\hypersetup{colorlinks=true,linkcolor=blue,urlcolor=blue,citecolor=blue}
\usepackage[cal=boondox]{mathalfa}
\usepackage{lipsum}
\usepackage{relsize}
\usepackage{color}
\usepackage{nameref}

\usepackage{soul}

\begin{document}
	
\title{Trace-free Einstein gravity as a constrained bigravity theory}
		
\author{Merced Montesinos}
\email{merced.montesinos@cinvestav.mx}

\affiliation{Departamento de F\'{i}sica, Cinvestav, Avenida Instituto Polit\'{e}cnico Nacional 2508, San Pedro Zacatenco,\\
07360 Gustavo A. Madero, Ciudad de M\'exico, M\'exico}

\author{Diego Gonzalez}
\email{dgonzalezv@ipn.mx}

\affiliation{Departamento de F\'{i}sica, Cinvestav, Avenida Instituto Polit\'{e}cnico Nacional 2508, San Pedro Zacatenco,\\
07360 Gustavo A. Madero, Ciudad de M\'exico, M\'exico}
\affiliation{Escuela Superior de Ingenier\'ia Mec\'anica y El\'ectrica, Instituto Polit\'ecnico Nacional, Unidad Profesional Adolfo L\'opez Mateos, Zacatenco, 07738 Gustavo A. Madero, Ciudad de M\'exico, M\'exico}
 
\date{\today}
	
\begin{abstract}
Trace-free Einstein gravity is a prominent alternative to general relativity, which has two versions: one in which the energy-momentum conservation is assumed a priori and another in which it is not. In the first version, the cosmological constant arises as an integration constant. We report two diffeomorphism-invariant actions for trace-free Einstein gravity in which the energy-momentum conservation emerges directly from the equations of motion. The first action is based on the metric, while the second action is based on the tetrad and connection where the conservation of the energy-momentum holds if torsion vanishes. To achieve this, auxiliary fields but otherwise dynamical are introduced in the corresponding actions, which have the form a constrained bigravity theory. These results constitute a significant advance in our understanding of trace-free Einstein gravity. 
\end{abstract}

\maketitle

\section{Introduction}

While being our best current theory of gravity, general relativity has some quite fundamental issues, such as the cosmological constant problem~\cite{Ellis_2011,Ellis_2014}, which have motivated the quest for alternative theories of gravity. Among the various alternative theories available, trace-free Einstein gravity stands out as an attractive theory because it is very closely related to general relativity. Trace-free Einstein gravity is an old theory~\cite{Einstein_1919,Einstein_1952,Einstein_1927}, almost as old as general relativity, in which only the trace-free part of Einstein’s equations for general relativity is adopted as the fundamental equations for the description of the gravitational phenomena, and so the cosmological constant does not appear in the equations of motion. In trace-free Einstein gravity, the energy-momentum conservation is no longer a consequence of the second Bianchi identity, as it happens in general relativity. Because of this, there are two versions of trace-free Einstein gravity. It must be stressed that both versions of the theory are diffeomorphism covariant (see Sec.~\ref{section_metric}). The first one works under the assumption of the energy-momentum conservation, whereas the second one allows for the nonconservation of energy-momentum. In the latter, the nonconservation of energy-momentum has nothing to do with torsion because the formulation involves the Levi-Civita connection. It is noteworthy that in the first version, as a consequence of the energy-momentum conservation and the second Bianchi identity, the cosmological constant naturally emerges as an integration constant~\cite{Ellis_2011,MontGonz_2023}. It is worth remarking that in trace-free Einstein gravity, as in general relativity, there is no unimodular condition of any kind. Up to now, the only diffeomorphism-invariant actions for trace-free Einstein gravity have been given in vacuum and within the context of (constrained) $BF$ theories~\cite{MontGonz_2023}.

In this paper, we go one step further and provide two formulations for trace-free Einstein gravity, one in the metric formalism and another in the tetrad and connection formalism, with the crucial feature that the energy-momentum conservation is not assumed a {\it priori} but instead emerges naturally from the equations of motion. In the metric formalism, we provide a totally diffeomorphism-invariant action that yields the trace-free Einstein equations. In this formulation, the cosmological constant arises as an integration constant, and this fact together with the second Bianchi identity imply the energy-momentum conservation. To accomplish this, the action involves not only the physical metric to which matter fields are coupled but also an auxiliary metric, which is dynamical, however. In this way, our formulation expresses trace-free Einstein gravity in the form of a constrained bigravity theory, which is distinct from the usual bigravity theory considered in the literature where no constraints are imposed~\cite{Hassan2012,deRham2014,deRham2015,KennaAllison2019}. On the other hand, in the tetrad and connection formalism, we also present a fully diffeomorphism-invariant action that leads to the trace-free Einstein equations. In this formulation, the cosmological constant also emerges naturally as an integration constant implied by the second Bianchi identity. This fact and the second Bianchi identity imply the energy-momentum conservation only if torsion vanishes. As in the metric formalism, the action also involves auxiliary fields, which are dynamical, however. More precisely, in addition to the physical tetrad and connection to which matter fields are coupled, the action depends on an auxiliary tetrad and connection. Another relevant aspect of the actions reported in this paper is the fact that they represent the first diffeomorphism-invariant actions for trace-free Einstein gravity, both in the metric formalism as well as in the tetrad and connection formalism.

\section{The action in the metric formalism}\label{section_metric}
In the metric formalism, trace-free Einstein gravity in four dimensions is given by the following equations of motion~\cite{Einstein_1927}:
\begin{equation}\label{TFEM}
R_{\mu\nu} - \frac14 R g_{\mu\nu} =\frac{8 \pi G}{c^4} \Big( T_{\mu\nu} - \frac14 T g_{\mu\nu}\Big),
\end{equation} 
where $R_{\mu\nu}=R^{\theta}{}_{\mu\theta\nu}$ is the Ricci tensor and $R=g^{\mu\nu} R_{\mu\nu}=R^{\mu\nu}{}_{\mu\nu}$ is the scalar curvature calculated with the components of the Riemman tensor coming from the Levi-Civita connection compatible with the metric $g_{\mu\nu}$, whereas $T_{\mu\nu}$ is the energy-momentum tensor. In addition to~\eqref{TFEM}, we have the equations for the matter fields. 

There are two diffeomorphism-covariant versions of the theory~\cite{MontGonz_2023}: (1) In the first version, the energy-momentum conservation is assumed, i.e., $\nabla_{\mu} T^{\mu\nu}=0$. This fact together with the second Biancli identity imply that the differential of $R + \frac{8 \pi G}{c^4} T$ vanishes,
\begin{equation}\label{d1}
d \left ( R + \frac{8 \pi G}{c^4} T \right ) =0,
\end{equation}
from which the cosmological constant $\Lambda$ emerges as an integration constant
\begin{equation}\label{Lambda_IC}
R + \frac{8 \pi G}{c^4} T =: 4 \Lambda. 
\end{equation}
(2) In the second version, the energy-momentum conservation is not assumed and the second Bianchi identity, instead of~\eqref{d1}, implies
\begin{equation}
\frac14 d \left( R + \frac{8 \pi G}{c^4} T \right) =\frac{8 \pi G}{c^4} \left( \nabla_{\mu} T^{\mu}{}_{\nu}\right) d x^{\nu}.  \label{dR_no_EC}
\end{equation}
Therefore, there is a major difference between any of the two versions of trace-free Einstein gravity and general relativity, in which energy-momentum conservation is not assumed, but implied precisely by the second Bianchi identity. Additionally, it is also important to remark that both theories, trace-free Einstein gravity and general relativity, do not involve any unimodular condition, which is a property of unimodular gravity~\cite{Anderson_1971,Weinberg_1989}. 

Thus, it would be desirable to have an action that naturally leads to trace-free Einstein gravity in the version described in item (1) above. This is, in fact, one of the main contributions of this paper. More precisely, we report an action that has the following properties: (a) It gives the equations of motion~\eqref{TFEM}. (b) It is diffeomorphism invariant. (c) The cosmological constant $\Lambda$ emerges as an integration constant. (d) The energy-momentum conservation is not assumed, but it is naturally deduced. This is, as far as we know, the first approach where the energy-momentum conservation is deduced. The price we have to pay for all of this is the introduction of an auxiliary metric $f_{\mu\nu}$, and this is the reason bigravity theory is involved in our approach. Nevertheless, the way in which bigravity theory is considered here is slightly different from the way it is usually considered in literature~\cite{Hassan2012,deRham2014,deRham2015,KennaAllison2019}. To be precise, we consider a constrained bigravity theory.     

Once the features of the action have been established, it is time to introduce it. The diffeomorphism-invariant action is given by
\begin{eqnarray}\label{action_bigravity}
S &=& \int_{{\cal M}^4} \left [ \frac{1}{2 \kappa^2} \sqrt{-g} R + \sqrt{-g} {\mathcal L}_{\rm matter}\right. \nonumber\\
&& \left. + \frac{1}{2\alpha^2} \sqrt{-f} R[f] - \frac{\lambda}{\kappa^2} \left ( \sqrt{-g} - \sqrt{-f}  \right ) \right ] d^4 x.
\end{eqnarray}
To avoid a cumbersome notation, here $g_{\mu\nu}$ is the physical metric and $g$ its determinant, $g=\det{(g_{\mu\nu})}$. $R = g^{\mu\nu} R_{\mu\nu} = R^{\mu\nu}{}_{\mu\nu}$ is the scalar curvature calculated with the components of the Riemann tensor coming from the Levi-Civita connection compatible with the metric $g_{\mu\nu}$. The second dynamical metric field is $f_{\mu\nu}$ and $f$ its determinant, $f=\det{(f_{\mu\nu})}$. $R[f]= f^{\mu\nu} R_{\mu\nu} [f]= R^{\mu\nu}{}_{\mu\nu} [f]$ is the scalar curvature calculated with the components of the Riemann tensor coming from the Levi-Civita connection compatible with the metric $f_{\mu\nu}$. The matter fields, denoted generically as $\psi$, couple only to the physical metric $g_{\mu\nu}$ in the usual way through ${\mathcal L}_{\rm matter}$; $\alpha^2$ and $\kappa^2$ are coupling constants. Last, and very important, $\lambda$ is a Lagrange multiplier imposing the constraint $\sqrt{-g} = \sqrt{-f}$. Note that the action~\eqref{action_bigravity} is a functional of $g_{\mu\nu}$, $f_{\mu\nu}$, $\lambda$, and $\psi$ and, by definition, all of these fields are dynamical in the sense that the action is varied with respect to them, which gives an equation of motion corresponding to each field, as follows:
\begin{eqnarray}
&& \delta g_{\mu\nu}: R_{\mu\nu} -\frac12 R g_{\mu\nu} + \lambda g_{\mu\nu} = \kappa^2 T_{\mu\nu}, \label{e1}\\
&& \delta f_{\mu\nu}: R_{\mu\nu} [f] - \frac12 R[f] f_{\mu\nu} -\frac{\alpha^2} {\kappa^2} \lambda  f_{\mu\nu} =0, \label{e2}\\
&& \delta\lambda: \sqrt{-g} = \sqrt{-f}, \label{e3}\\
&& \delta \psi: \mbox{matter field equations}, \label{e4}
\end{eqnarray}
where the energy-momentum tensor $T_{\mu\nu}$ is defined by $\delta \int_{{\cal M}^4} \sqrt{-g} {\mathcal L}_{\rm matter} d^4 x =\frac12 \int_{{\cal M}^4} T^{\mu\nu} \delta g_{\mu\nu} \sqrt{-g} d^4x$. 

The equation of motion~\eqref{e2} and the second Bianchi identity (that comes from the Levi-Civita connection constructed with metric $f_{\mu\nu}$) imply
\begin{equation}\label{derivada}
d \lambda =0.
\end{equation}
This result and the equation of motion~\eqref{e1} (together with the second Bianchi identity that comes from the Levi-Civita connection constructed with the metric $g_{\mu\nu}$) imply that the energy-momentum is conserved,
\begin{equation}
\nabla_{\mu} T^{\mu\nu}=0.    
\end{equation}
Furthermore, contracting the equation of motion~\eqref{e1} with $g^{\mu\nu}$, we get
\begin{equation}\label{first_trace}
4 \lambda = R + \kappa^2 T.
\end{equation}
Substituting this expression for $\lambda$ in~\eqref{e1}, we obtain 
\begin{equation}\label{TFEM2}
R_{\mu\nu} - \frac14 R g_{\mu\nu} = \kappa^2 \Big( T_{\mu\nu} - \frac14 T g_{\mu\nu}\Big),
\end{equation} 
which are precisely the trace-free Einstein equations~\eqref{TFEM}. Furthermore, because of~\eqref{derivada} and~\eqref{first_trace}, the cosmological constant $\Lambda$ emerges as an integration constant
\begin{equation}\label{emerging_Lambda}
4 \lambda = R + \kappa^2 T =: 4 \Lambda.
\end{equation}
Finally, substituting $\lambda= \Lambda$ in~\eqref{e2}, we get
\begin{equation}
R_{\mu\nu} [f] - \frac12 R[f] f_{\mu\nu} - \frac{\alpha^2} {\kappa^2} \Lambda f_{\mu\nu} =0.
\end{equation}

Although a previous diffeomorphism-invariant formulation of trace-free Einstein gravity was reported in Ref.~\cite{MontGonz_2023}, the formulation~\eqref{action_bigravity} is the first of its kind to employ the metric field $g_{\mu\nu}$ as the fundamental variable, in full agreement with Einstein's proposal~\cite{Einstein_1927}. Moreover, the formulation~\eqref{action_bigravity} includes also matter fields, whereas the action of Ref.~\cite{MontGonz_2023} does not (up to now). Furthermore, the formulation~\eqref{action_bigravity} has also the remarkable property that it also implies the conservation of the energy-momentum, a fact that is not trivial and that seems not easy to achieve. The formulation~\eqref{action_bigravity} of trace-free Einstein gravity is the first that has this feature and is, in this sense, on the same footing as general relativity, where energy-momentum is also naturally conserved. In addition, it is important to emphasize that the cosmological constant $\Lambda$~\eqref{emerging_Lambda} emerges as an integration constant implied by the second Bianchi identity, i.e., as a consequence of the geometric content of the theory, in full agreement with Einstein's proposal. Finally, some words regarding the auxiliary metric $f_{\mu\nu}$. We expect that it plays the same role as in unconstrained bigravity theory~\cite{Hassan2012,deRham2014,deRham2015,KennaAllison2019}.

\section{The action in the tetrad and connection formalism}\label{section_tetrad}

For the sake of completeness, we now turn to the corresponding version of the action~\eqref{action_bigravity} in the first-order formalism. The action that is the subject of this section describes trace-free Einstein gravity as two copies of the Holst action supplemented with a constraint. Here, we also consider Lorentzian manifolds and the frame rotation group corresponds to $SO(3,1)$. Capital indices from the middle of the alphabet $I, J, K, \ldots$ are raised and lowered with the metric $(\eta_{IJ}) = \mbox{diag} (-1,1,1,1)$, and $\varepsilon_{IJKL}$ is the totally anti-symmetric $SO(3,1)$ invariant tensor with $\varepsilon_{0123}=1$. 

The diffeomorphism-invariant action is given by
\begin{align}\label{TC_action}
S&= \int_{{\cal M}^4} \left[\frac{1}{2\kappa^2} \left( \ast(e^I \wedge e^J) \wedge \mathcal{R}_{IJ} - \frac{1}{\gamma} e^I \wedge e^J \wedge \mathcal{R}_{IJ} \right)\right. \nonumber \\
& +  L_{\rm matter} + \frac{1}{2\alpha^2} \left( \ast(\theta^I \wedge \theta^J) \wedge \mathcal{F}_{IJ} - \frac{1}{\xi} \theta^I \wedge \theta^J \wedge \mathcal{F}_{IJ} \right) \nonumber\\ 
&\left.-\frac{\lambda}{\kappa^2} \left( \eta - \rho \right) \right] ,
\end{align}
where $e^I$ is an orthonormal frame of one-forms and  $\omega_{IJ}=-\omega_{JI}$ is a connection with curvature $\mathcal{R}^I{}_J = d \omega^I{}_J + \omega^I{}_K \wedge \omega^K{}_J$. Also, $\theta^I$ is an auxiliary orthonormal frame of one-forms, while $A_{IJ}=-A_{JI}$ is an auxiliary connection, which has curvature $\mathcal{F}^I{}_J = d A^I{}_J + A^I{}_K \wedge A^K{}_J$. The matter fields $\psi$ only couple to the frame $e^I$ and to the connection $\omega^I{}_J$ through $L_{\rm matter}$, $\alpha^2$ and $\kappa^2$ are coupling constants, and $\gamma$ and $\xi$ are nonvanishing real parameters. Furthermore, $\eta:=\frac{1}{4!} \varepsilon_{IJKL} e^I \wedge e^J \wedge e^K \wedge e^L$ and $\rho:=\frac{1}{4!} \varepsilon_{IJKL} \theta^I \wedge \theta^J \wedge \theta^K \wedge \theta^L$ are volume forms, and $\ast$ stands for the internal dual and so $\ast(e^I \wedge e^J)= \frac12 \varepsilon^{IJ}{}_{KL} e^K \wedge e^L$ and 
$\ast(\theta^I \wedge \theta^J)= \frac12 \varepsilon^{IJ}{}_{KL} \theta^K \wedge \theta^L$. Last, $\lambda$ is a Lagrange multiplier imposing the constraint $\eta = \rho$. Note that the action~\eqref{TC_action} is a functional of $e^I$, $\theta^I$, $\omega_{IJ}$, $A_{IJ}$, $\lambda$, and $\psi$, and thus all of these fields are dynamical in the sense that the action is going to be varied with respect to them. The equations of motion coming from the variation of \eqref{TC_action} with respect to these fields are
\begin{align}
    \delta e^I \,:&\,\, \varepsilon_{IJKL} e^J \wedge \mathcal{R}^{KL} - \frac{2}{\gamma} e^J \wedge \mathcal{R}_{IJ} \nonumber\\
    &\,\,-\frac{\lambda}{3} \varepsilon_{IJKL} e^J \wedge e^K \wedge e^L +2 \kappa^2 T_{JI} \star e^J=0, \label{EOM_tet_e} \\
    \delta \omega^{IJ} \,&:\,\, D_{\omega} e^I=T^I,  \label{EOM_the_omega}\\
    \delta \theta^I \,:&\,\, \varepsilon_{IJKL} \theta^J \wedge \mathcal{F}^{KL} - \frac{2}{\xi} \theta^J \wedge \mathcal{F}_{IJ} \nonumber\\
    &\,\,+\frac{\alpha^2 \mu}{3  \kappa^2} \varepsilon_{IJKL} \theta^J \wedge \theta^K \wedge \theta^L=0, \label{EOM_tet_theta} \\
    \delta A^{IJ} \,&:\,\, D_A \theta^I=0,  \label{EOM_the_A}\\
    \delta \lambda \,&:\,\, \eta = \rho, \label{EOM_the_mu}\\
    \delta \psi \,&:\,\, \mbox{matter field equations},
\end{align}
where $D_\omega$ and $D_A$ are the covariant derivatives defined by $\omega^I{}_J$ and $A^I{}_J$, respectively. Also, $\star$ denotes the Hodge dual and then $\star e_I=\frac{1}{3!} \varepsilon_{IJKL} e^J \wedge e^K \wedge e^L$. The energy-momentum tensor $T_{IJ}$ is defined by $\delta L_{matter}=\delta e^I \wedge T_{JI}  \allowbreak \star e^J$. Note that $T_{IJ}$ is asymmetric in general (see Ref.~\cite{Montesinos_2018} for specific expressions of $T_{IJ}$). In addition, $T^I$ is the torsion of $\omega^I{}_J$, which might be induced by the couplings of the matter fields. In particular, it is nonvanishing for fermion fields~\cite{Romero_2021,Romero_2022}. 

The equation of motion~\eqref{EOM_the_A} means that $A^I{}_J$ is a torsion-free spin connection. Then, because of the corresponding first Bianchi identity $\mathcal{F}^I{}_J \wedge \theta^J =0$, the term involving the parameter $\xi$ in \eqref{EOM_tet_theta} vanishes. Taking into account this and writing the curvature ${\mathcal R}_{IJ}$ in the basis $e^I$ as ${\mathcal R}_{IJ} = \frac12 R_{IJKL} e^K \wedge e^L$ and the curvature ${\mathcal F}_{IJ}$ in the basis $\theta^I$ as ${\mathcal F}_{IJ} = \frac12 F_{IJKL} \theta^K \wedge \theta^L$, the equations of motion \eqref{EOM_tet_e} and \eqref{EOM_tet_theta} read
\begin{align}
    &R_{IJ}-\frac12 R \eta_{IJ} + \frac{1}{2 \gamma} \varepsilon^{KLM}{}_{I} R_{JKLM} + \lambda \eta_{IJ} = \kappa^2 T_{IJ}, \label{tet_eqR}\\
    &F_{IJ}-\frac12 F \eta_{IJ} - \frac{\alpha^2\lambda}{\kappa^2} \eta_{IJ} =0, \label{tet_eqF}
\end{align}
respectively. In \eqref{tet_eqR}, $R_{IJ}=R^K{}_{IKJ}$ are the components of the Ricci tensor and $R=R^{IJ}{}_{IJ}$ is the scalar curvature. Note that $R_{IJ}$ is nonsymmetric if there is nonvanishing torsion. Analogously, in \eqref{tet_eqF}, $F_{IJ}=R^K{}_{IKJ}$ are the components of the Ricci tensor and $F=F^{IJ}{}_{IJ}$ is the scalar curvature. Notice that \eqref{tet_eqR} and \eqref{tet_eqF} are the corresponding versions of~\eqref{e1} and~\eqref{e2} in the orthonormal frames $e^I$ and $\theta^I$. Notice also that because of the Bianchi identity $\mathcal{R}^I{}_J \wedge e^J =D_\omega T^I$, the term involving the parameter $\gamma$ in \eqref{tet_eqR} does not vanish if there is nonvanishing torsion.

Following a procedure completely analogous to that presented in Sec.~\ref{section_metric}, we note that~\eqref{tet_eqF} and the second Bianchi identity (that comes from the connection $A^I{}_J$) imply 
\begin{eqnarray}\label{key_missing}
d\lambda=0,
\end{eqnarray}
which along with \eqref{tet_eqR} and the second Bianchi identity (that comes from the connection $\omega^I{}_J$) imply 
\begin{align}\label{tet_conserv}
    \nabla_I T^{IJ}=& \frac{1}{\kappa^2} \bigg( \frac12 R^{KLIJ} T_{IKL} - R_{IK} T^{KIJ} \nonumber \\
    & + \frac{1}{2\gamma} R^{JIKL} T_K{}^{MN} \varepsilon_{MNLI} \bigg),
\end{align}
where $\nabla_K T^I{}_J$ is defined by $D_\omega T^I{}_J=\left(\nabla_K T^I{}_J \right) e^K$, and  $T^I{}_{JK}$ is defined by $T^I= \frac12 T^{I}{}_{JK} e^J \wedge e^K$. From~\eqref{tet_conserv} it is clear that the energy-momentum is not conserved if there is nonvanishing torsion. If torsion vanishes, the energy-momentum conservation arises naturally.  

Contracting \eqref{tet_eqR} with $\eta^{IJ}$, we arrive at $4 \lambda = R + \kappa^2 T + \frac{1}{2\gamma} R_{IJKL} \varepsilon^{IJKL} $, and substituting this result in \eqref{tet_eqR}, we obtain
\begin{align}
    &R_{IJ}-\frac14 R \eta_{IJ} - \frac{1}{2 \gamma} \bigg( \varepsilon_{I}{}^{KLM} R_{JKLM} \nonumber\\ 
    &- \frac14 R_{KLMN} \varepsilon^{KLMN} \eta_{IJ} \bigg)   = \kappa^2 \bigg( T_{IJ}-\frac14 T \eta_{IJ} \bigg).
\end{align}
Clearly, the cosmological constant $\Lambda$ emerges as an integration constant implied by~\eqref{key_missing} and $4 \lambda = R + \kappa^2 T + \frac{1}{2\gamma} R_{IJKL} \varepsilon^{IJKL}$, no matter if the energy-momentum is conserved or not
\begin{equation}\label{tet_cosmo}
      4 \lambda = R + \kappa^2 T + \frac{1}{2\gamma} R_{IJKL} \varepsilon^{IJKL}=:4\Lambda.
\end{equation}
Furthermore, substituting $\lambda=\Lambda$ in~\eqref{tet_eqF}, it acquires the form
\begin{equation}
    F_{IJ}-\frac12 F \eta_{IJ} - \frac{\alpha^2}{\kappa^2} \Lambda \eta_{IJ} =0. \label{tet_eqF2}
\end{equation}
The formulation of trace-free Einstein gravity given in this section has some relevant features: (i) The action~\eqref{TC_action} is the first diffeomorphism-invariant action of the theory reported in literature that is given in terms of the tetrad and connection. (ii)~The cosmological constant $\Lambda$ emerges naturally in the formulation as an integration constant, which is implied by the second Bianchi identity. (iii) The conservation of the energy-momentum also arises naturally if torsion vanishes as in the metric formalism reported in Sec.~\ref{section_metric}. However, if there is nonvanishing torsion, the right-hand side of~\eqref{tet_conserv} does not vanish in the general case.

\section{Conclusions}\label{section_conclusions}
In this paper, we have posed two formulations of trace-free Einstein gravity in which the energy-momentum conservation is deduced naturally from the equations of motion. In the first formulation, we have described trace-free Einstein gravity in the metric formalism through an action that depends on two dynamical metrics: the physical metric $g_{\mu\nu}$ that couples to matter fields and an auxiliary metric $f_{\mu\nu}$. In this formulation, the conservation of energy-momentum is a consequence of the fact that the cosmological constant emerges as an integration constant. In the second formulation, we have done so in the tetrad and connection formalism using an action that depends on two sets of dynamical tetrads and connections: the physical tetrad and connection that couple to matter fields and an auxiliary tetrad and connection. More precisely, the action describes trace-free Einstein gravity as two copies of the Holst action supplemented with a constraint on the volume forms. Note that in this formulation the cosmological constant arises as an integration constant no matter if the energy-momentum is conserved or not. Moreover, the energy-momentum conservation only holds if torsion vanishes. Furthermore, in both formulations, the (effective) cosmological constant for the auxiliary gravitational sector is modulated by the coupling constants $\alpha^2$ and $\kappa^2$ and has an opposite sign compared with the one of the physical gravitational sector. We hope that the unexpected relation between trace-free Einstein gravity and bigravity theory reported in this paper motivates the research in both theories. We would also like to mention that it has been speculated in Ref.~\cite{PhysRevLett.122.221302} that massive fermions should lead to violation of the energy-momentum conservation. Therefore, it is remarkable that we have reached in this paper the same conclusion from first principles in the context of the tetrad and connection formulation of trace-free Einstein gravity~\eqref{TC_action}. However, our analysis is not restricted to fermions, but to all matter fields that have nonvanishing torsion.

On the other hand, it is worth recalling that a tetrad and connection formulation of trace-free Einstein gravity with matter fields was previously given in Ref.~\cite{MontGonz_2023}. Nevertheless, there is no action leading to the equation of motion of such a formulation. In addition, there are two versions of it, one in which the energy-momentum conservation is assumed and one in which it is not. So, the formulation given in this paper is different from that of Ref.~\cite{MontGonz_2023}.

The current formulations of trace-free Einstein gravity can also be regarded as generalizations of unimodular gravity~\cite{Anderson_1971,Weinberg_1989}. This interpretation is clear from the fact that, unlike unimodular gravity where the volume is fixed, the volume in these formulations is dynamical. However, they offer much more than that. These are the first formulations of trace-free Einstein gravity where the energy-momentum conservation is deduced, not assumed. Clearly, these facts are absent in unimodular gravity. 

An important aspect of the formulation~\eqref{action_bigravity} is that, although it was presented in four dimensions, it also holds in $n$ dimensions. Similarly, the action~\eqref{TC_action} can also be generalized to $n$-dimensions but instead of considering two copies of the Holst action, we must consider just two copies of the Palatini action. Additionally, we can also obtain a second-order formulation of the theory by eliminating $\omega^I{}_J$ and $A^I{}_J$ from the action~\eqref{TC_action} using their equations of motion. The resulting action would depend functionally on the tetrads (or vielbein fields in $n$ dimensions) $e^I$ and $\theta^I$ and the matter fields.  

Finally, it would be very valuable to perform the Hamiltonian analyses of these formulations, which is crucial for the canonical quantization of the theory~\cite{RovBook,ThieBook}. For the first-order action~\eqref{TC_action}, such an analysis can be performed using the parametrization of the frame and the connection used in Refs.~\cite{PhysRevD.101.024042,PhysRevD.101.084003} (see also~\cite{ Romero_2021,Romero_2022}). In this context, it would also be relevant to analyze the role of the parameters $\gamma$ and $\xi$ involved in the formulation~\eqref{TC_action}.

\acknowledgments
We thank Mariano Celada, Alejandro Perez, and Ricardo Escobedo for fruitful discussions on the subject. D. G. acknowledges the financial support of Instituto Polit\'ecnico Nacional, Grant No. SIP-20240184, and the postdoctoral fellowship from Consejo Nacional de Humanidades, Ciencia y Tecnología (CONAHCyT), México.


\bibliography{references}
	
\end{document}